\documentclass[%
 aip,
 amsmath,amssymb,
 reprint,
]{revtex4-1}

\usepackage{graphicx}
\usepackage[caption=false]{subfig}
\usepackage{epstopdf}

\begin{document}

\title[Single Gate P-N Junctions in Graphene-Ferroelectric Devices]{Single Gate P-N Junctions in Graphene-Ferroelectric Devices}

\author{J. Henry Hinnefeld}
 \affiliation{Department of Physics, University of Illinois at Urbana-Champaign, Urbana, Illinois 61801, USA}
\author{Ruijuan Xu}
 \affiliation{Department of Materials Science and Engineering, University of California, Berkeley, California 94720, USA}
 \affiliation{Materials Science Division, Lawrence Berkeley National Laboratory, Berkeley, CA 94720, USA}
\author{Steven Rogers}
 \affiliation{Department of Materials Science and Engineering, University of Illinois at Urbana-Champaign, Urbana, Illinois 61801, USA}
\author{Shishir Pandya}
 \affiliation{Department of Materials Science and Engineering, University of California, Berkeley, California 94720, USA}
 \affiliation{Materials Science Division, Lawrence Berkeley National Laboratory, Berkeley, CA 94720, USA}
\author{Moonsub Shim}
 \affiliation{Department of Materials Science and Engineering, University of Illinois at Urbana-Champaign, Urbana, Illinois 61801, USA}
\author{Lane W. Martin}
 \affiliation{Department of Materials Science and Engineering, University of California, Berkeley, California 94720, USA}
 \affiliation{Materials Science Division, Lawrence Berkeley National Laboratory, Berkeley, CA 94720, USA}
\author{Nadya Mason}
 \affiliation{Department of Physics, University of Illinois at Urbana-Champaign, Urbana, Illinois 61801, USA}
 \email{nadya@illinois.edu}

\date{\today}

\keywords{graphene, p-n junction, ferroelectrics, PZT}

\begin{abstract}
Graphene's linear dispersion relation and the attendant implications for bipolar electronics applications have motivated a range of experimental efforts aimed at producing p-n junctions in graphene. Here we report electrical transport measurements of graphene p-n junctions formed via simple modifications to a PbZr$_{0.2}$Ti$_{0.8}$O$_3$ substrate, combined with a self-assembled layer of ambient environmental dopants. We show that the substrate configuration controls the local doping region, and that the p-n junction behavior can be controlled with a single gate. Finally, we show that the ferroelectric substrate induces a hysteresis in the environmental doping which can be utilized to activate and deactivate the doping, yielding an `on-demand' p-n junction in graphene controlled by a single, universal backgate.
\end{abstract}

\maketitle

Graphene is a subject of intense research interest due to the enormous potential of its electronic and mechanical properties\cite{Geim2007}. In particular, p-n junctions in graphene have great potential for both fundamental research and commercial applications, and have been utilized to study the quantum Hall effect \cite{Williams2007, Ozyilmaz2007, Velasco2010} and Klein tunneling\cite{Stander2009, Young2009} as well as to fabricate flexible transistors\cite{Kim2010}. Previous work on p-n junctions in graphene employed multiple electrostatic gates \cite{Meric2008, Williams2007, Ozyilmaz2007, Huard2007, Liu2008, Stander2009, Velasco2009, Velasco2010, Young2009}, charge transfer from the controlled deposition of chemical adsorbates \cite{Farmer2009,Lohmann2009,Brenner2010,Cheng2011,Sojoudi2012,Seo2014,Park2015}, high current-induced charging of trap states in the substrate \cite{Chiu2010}, or periodically poled ferroelectric substrates\cite{Baeumer2015}.

In this Letter we report the fabrication of p-n junctions in graphene deposited on a uniformly poled ferroelectric substrate. The local doping in our devices is accomplished by combining simple modifications to a lead zirconium titanate (PbZr$_{0.2}$Ti$_{0.8}$O$_3$) substrate -- the evaporation of thin SiO$_2$ films in some regions -- with a self-assembled layer of environmental dopants. We find that the PbZr$_{0.2}$Ti$_{0.8}$O$_3$ substrate modulates the doping effect of the adsorbed dopants: devices are exposed to ambient conditions after fabrication whereupon experimental observations confirm both the presence of adsorbed dopants (likely primarily H$_2$O) and their enhanced doping effect on the PbZr$_{0.2}$Ti$_{0.8}$O$_3$ relative to the SiO$_2$.
Furthermore, we demonstrate that the PbZr$_{0.2}$Ti$_{0.8}$O$_3$ substrate induces a hysteresis in the environmental doping which can be used to activate and deactivate the doping via the application of large gate voltages. We employ this effect to create p-n junctions which can be reversibly transitioned between p-n junction and uniformly conducting configurations.

\begin{figure}
  \captionsetup[subfigure]{labelformat=empty}
  \subfloat[]{\includegraphics[width=.14\textwidth]{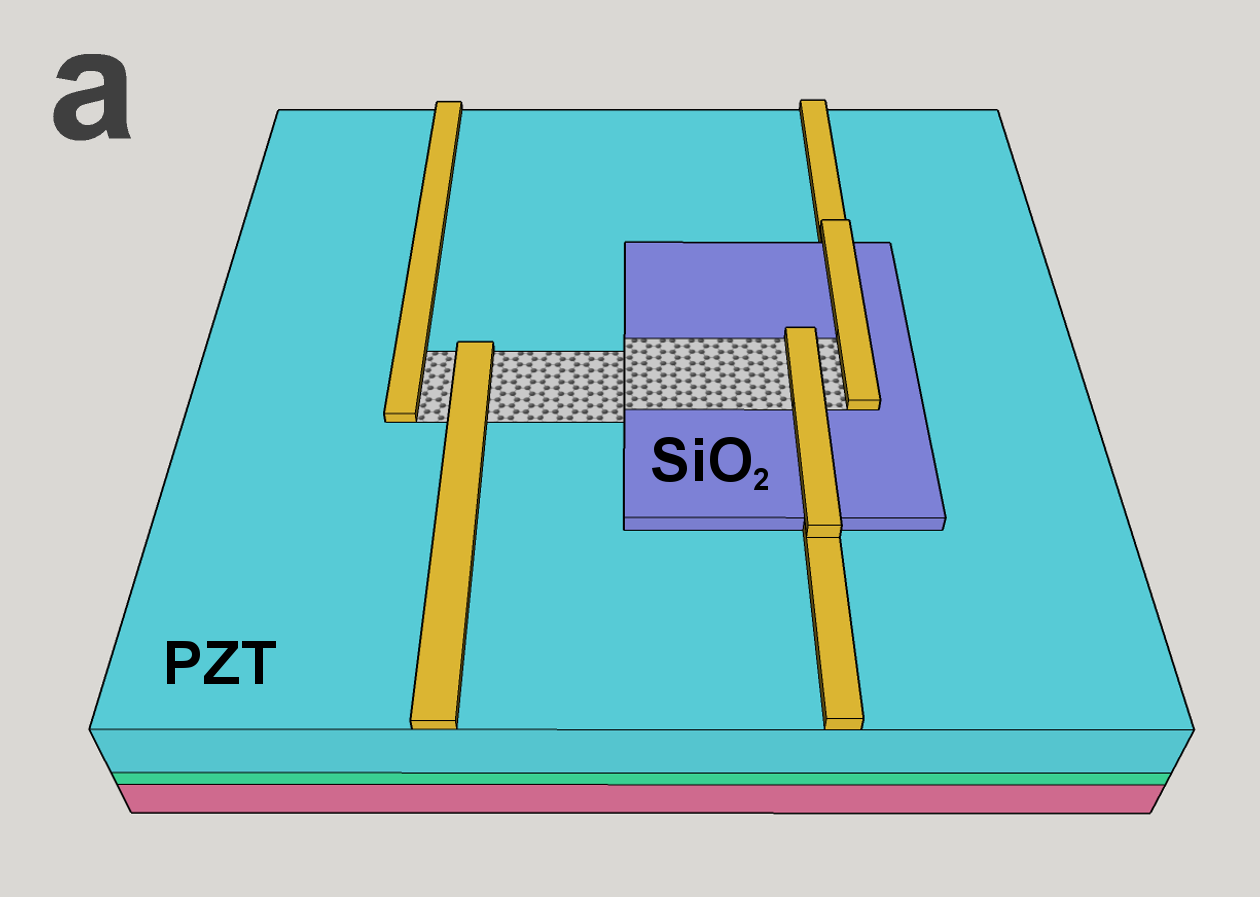}}\quad
  \subfloat[]{\includegraphics[width=.14\textwidth]{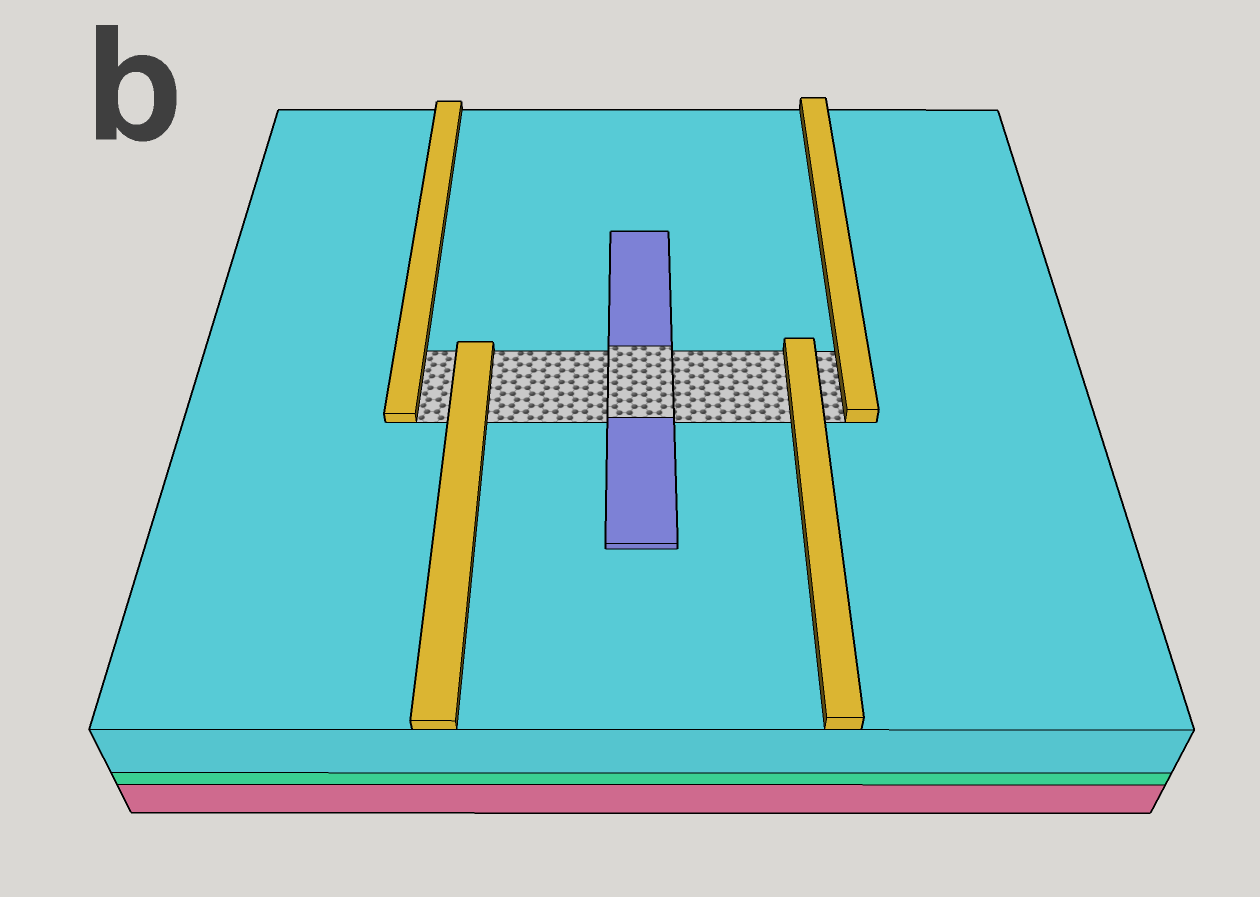}}\quad
  \subfloat[]{\includegraphics[width=.14\textwidth]{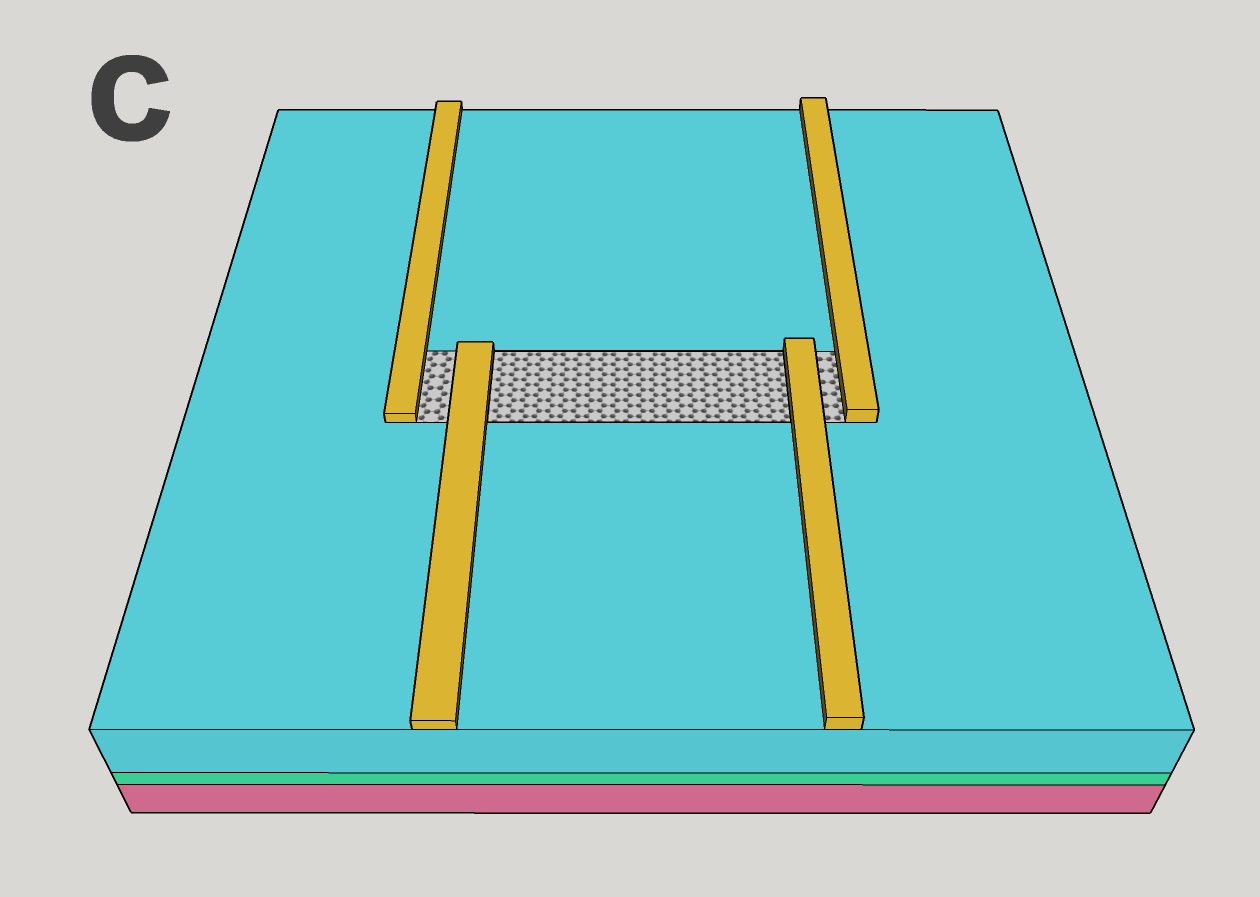}}\quad
  \subfloat[]{\includegraphics[width=.45\textwidth]{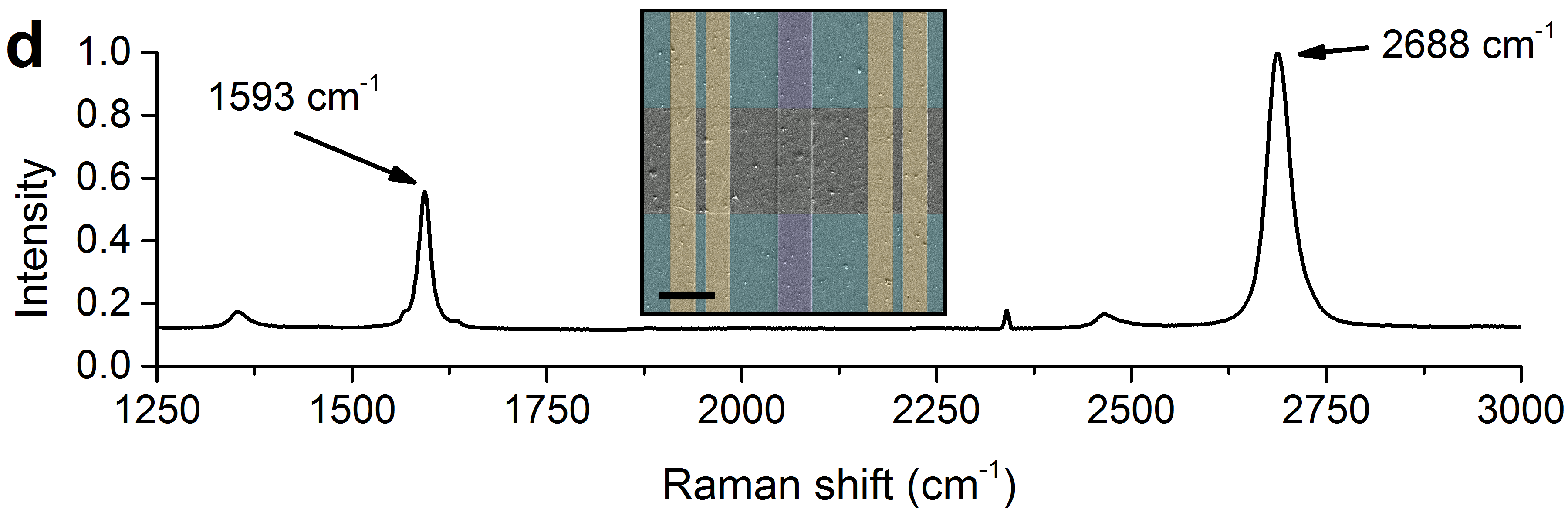}}
  \caption{
   (a-c) Schematic illustration of the device geometry. The graphene (grey) spans the regions of evaporated SiO$_2$ (purple) on the bare substrate (teal) and is contacted by Cr/Au electrodes (yellow). The universal backgate is shown in red. The evaporated SiO$_2$ region width ranges from 0.5 $\mu$m to 3 $\mu$m in the measured devices. The graphene channel is 6 $\mu$m by 4 $\mu$m measured from the inner contacts.
   (d) Raman spectra of graphene from a device fabricated on a thermal SiO$_2$ substrate. 
   Inset: a false-color SEM micrograph of a representative device. The scale bar is 2 $\mu$m.
  }
  \label{fig:devices}
\end{figure}

Devices consist of graphene micro-ribbons deposited on substrates which are partially covered by a thin layer of evaporated SiO$_2$, and contacted in a four-point geometry, as illustrated in Figures \ref{fig:devices}a-c. An SEM micrograph of a typical device is shown in the inset of Figure \ref{fig:devices}d. The devices are fabricated using standard lithography and deposition techniques on thin-film ferroelectric substrates. For the ferroelectric substrates, 120 nm thick (001)-oriented lead zirconium titanate (PbZr$_{0.2}$Ti$_{0.8}$O$_3$) films are prepared by pulsed-laser deposition (PLD) on a strontium titanate (SrTiO$_3$) substrate coated with 20 nm of strontium ruthenate (SrRuO$_3$), following established procedures\cite{Xu2014,Karthik2012}. For each substrate, an 80 nm-thick layer of SiO$_2$ is evaporated in small rectangular regions, with region widths ranging from 0.5 $\mu$m to 3 $\mu$m, as illustrated in Figs. \ref{fig:devices}a-c. CVD graphene is then transferred using standard wet transfer techniques\cite{Li2009}, and patterned into ribbons spanning the deposited SiO$_2$ using photolithography and reactive ion etching. The graphene channel is 6 $\mu$m by 4 $\mu$m measured from the inner contacts. Control devices which span regions with no evaporated SiO$_2$ are also fabricated. Finally, Cr/Au (3 nm/20 nm) leads are deposited in a four-point measurement configuration.  Raman spectroscopy is used to confirm the monolayer character and high quality of the graphene after fabrication; a representative spectra is shown in Figure \ref{fig:devices}d.

Transport measurements in air are performed using two Keithley 2400 SourceMeters. Measurements in vacuum are performed using an Agilent 4156C Semiconductor Parameter Analyzer. In both cases source-drain current is measured with a constant source-drain bias of 5 mV while the voltage applied to the backgate is swept. Gate leakage distorts transport results for gate voltages more positive than 1.5 V or more negative than -1 V, so gate voltages are limited to this range. Gate voltage sweep rates range from 10 mV/s to 100 mV/s; the data presented here are from sweeps at 100 mV/s. Slower sweep rates yield qualitatively similar results.

\begin{figure}[!b]
  \captionsetup[subfigure]{labelformat=empty}
  \subfloat[]{\includegraphics[width=.22\textwidth]{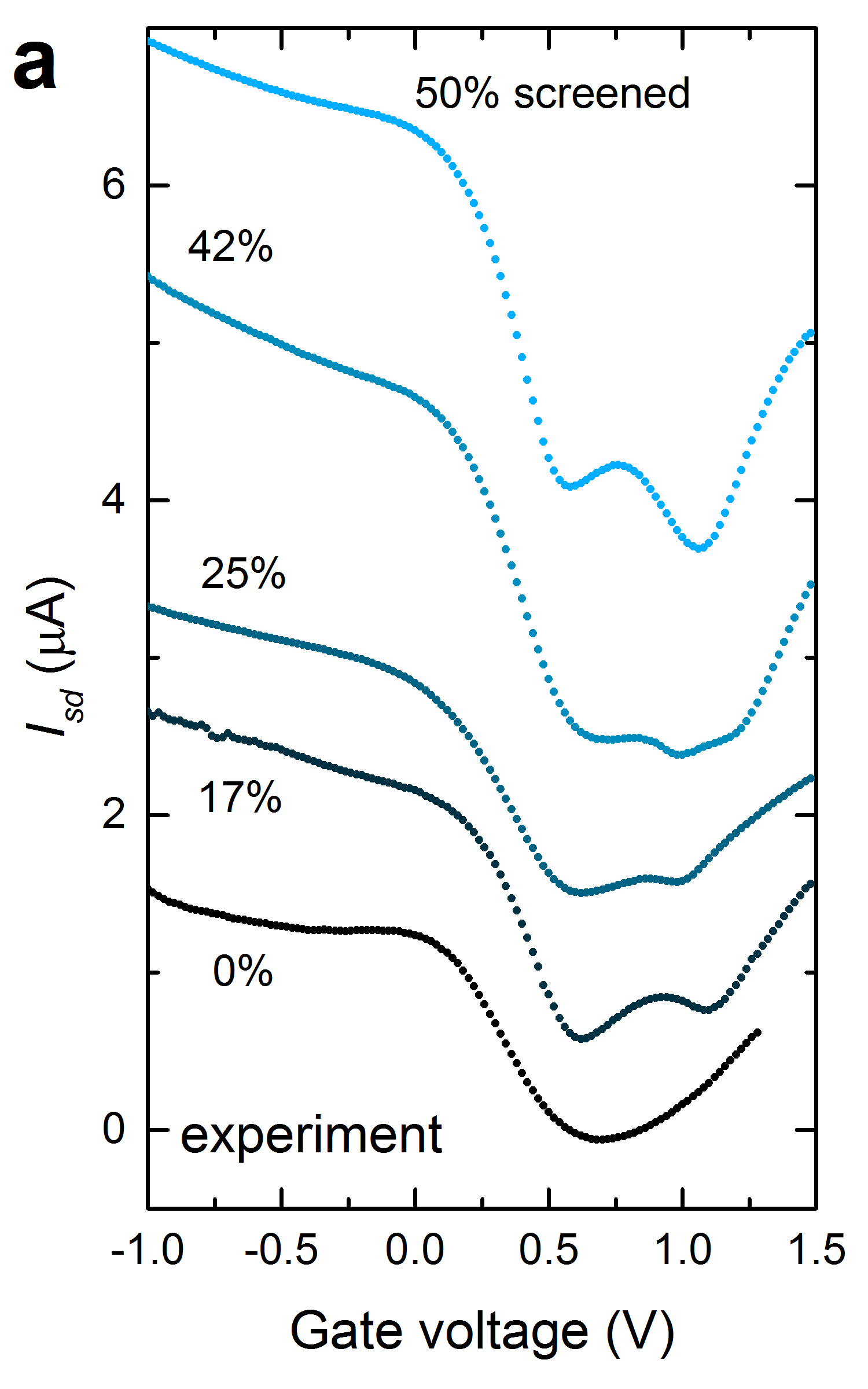}}\quad
  \subfloat[]{\includegraphics[width=.22\textwidth]{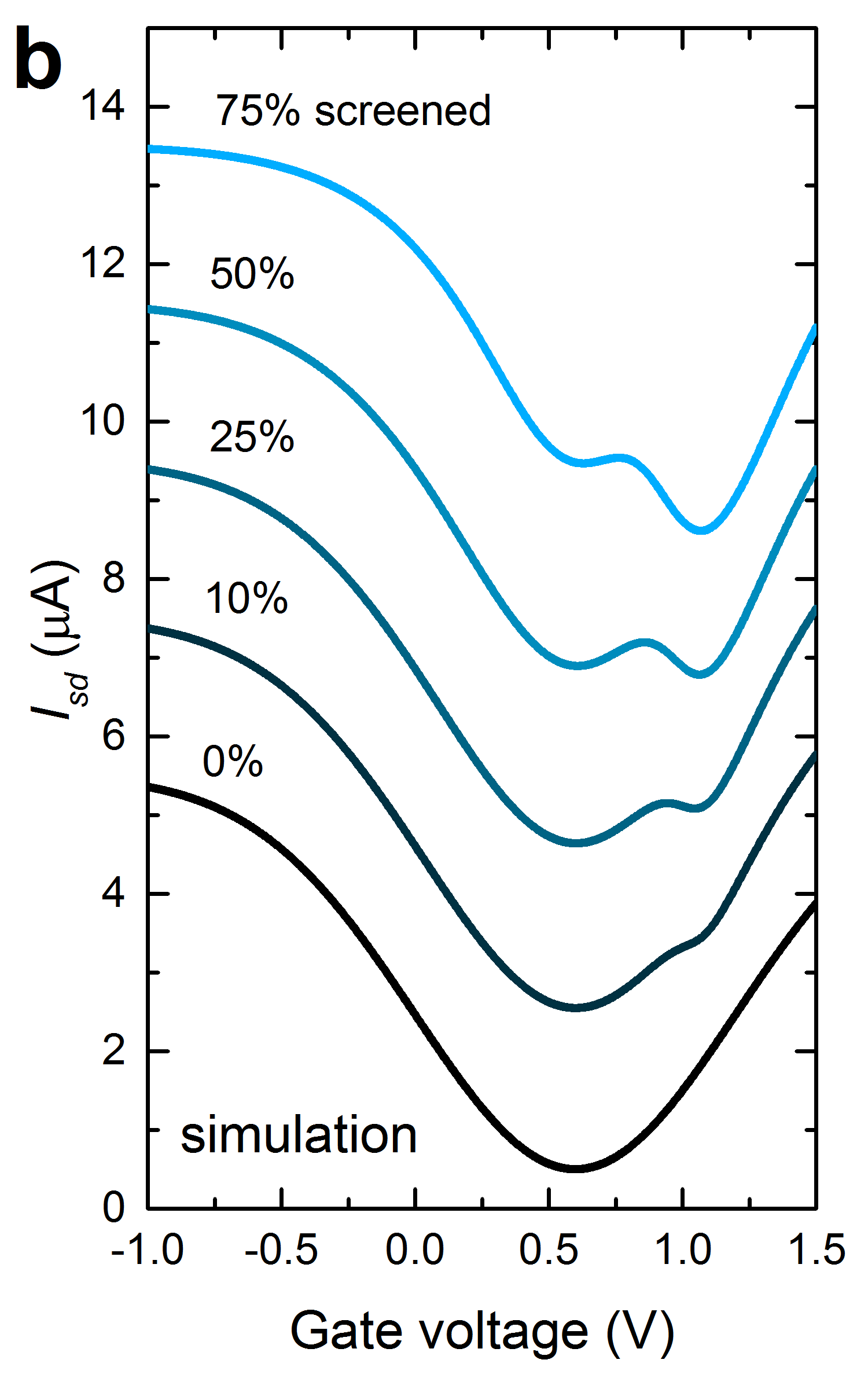}}\quad

  \caption{
   (a) Offset $I_{sd}$ vs $V_{gate}$ curves for devices having different widths of evaporated SiO$_2$ on a PbZr$_{0.2}$Ti$_{0.8}$O$_3$ substrate. The evaporated SiO$_2$ serves to locally screen the graphene from any substrate-dependent effects; from bottom to top the curves correspond to graphene channels which are 0, 17, 25, 42, and 50 percent screened. The devices having evaporated SiO$_2$ show split Dirac points, and the relative dominance of the left and right Dirac points can be tuned by varying the evaporated SiO$_2$ width.
   (b) Simulations of transport across a graphene device having two different locally doped regions, as a function of applied gate voltage. The simulations show both Dirac point splitting and variations in left vs. right Dirac point dominance as a function of screening region width, consistent with the experimental data. From bottom to top the curves correspond to graphene channels which are 0, 10, 25, 50, and 75 percent screened.
}
\label{fig:PZTdata}
\end{figure}

Figure \ref{fig:PZTdata}a shows room temperature $I_{sd}$ vs $V_{gate}$ curves for devices having different widths of evaporated SiO$_2$ on a PbZr$_{0.2}$Ti$_{0.8}$O$_3$ substrate. For the data shown here, the fraction of the graphene channel screened by evaporated SiO$_2$ ranges from 0\% to 50\% (corresponding to SiO$_2$ widths of 0 to 3 microns). Two features of the data are immediately apparent: first, for devices which span an evaporated SiO$_2$ region, the characteristic conductance minimum typically observed in graphene at the Dirac point is split into two distinct minima, one at the original Dirac point location and a second shifted to the right. This is apparent in the top and bottom curves of the Figure: the bottom curve, corresponding to a device with 0\% screening, displays a single minimum, while the top curve, corresponding to a device with 50\% screening, displays two pronounced minima. Second, the width of the evaporated SiO$_2$ region determines which of the two minima has a smaller absolute value. As the screening fraction is increased, `weight' is transferred from the minimum at the original Dirac point location to the secondary minimum, and the depths of the two minima vary accordingly. We attribute both of these effects to the presence of two different doping regions in the graphene, defined by the PbZr$_{0.2}$Ti$_{0.8}$O$_3$ substrate and the evaporated SiO$_2$.

The data can be understood by considering that as the gate voltage is swept from negative to positive, the Fermi level passes through the charge-neutrality point (CNP) of each graphene region separately. Taking the conductance to be linear with carrier density\cite{Hwang2007} $\sigma \propto k_F \langle\tau\rangle \propto n$ and the carrier density to be linear with the thermally smeared energy difference between the Fermi level and the CNP\cite{CastroNeto2009}, we model the conductance in the vicinity of the CNP as: $\rho^{-1} \propto n \propto 1 - e^{{-(V_{gate}}-\mu-\delta)^2/2c^2} + \epsilon$ where the constant $\mu$ accounts for the extrinsic doping introduced by the fabrication process, $\delta \in \{0,1\}$ describes the substrate-dependent doping, and $\epsilon$ accounts for the non-vanishing carrier density at the CNP. Assuming diffusive transport in the graphene, the relative weight of each separately doped region, and therefore the relative magnitude of the measured conductance minima, is determined by the fraction of the graphene channel which is screened:
$$I_{sd} \propto \left[ \rho_\text{scr.} \times (\text{pct.}_\text{scr.}) + \rho_\text{non-scr.} \times (\text{pct.}_\text{non-scr.}) \right]^{-1}$$
This is simulated in Figure \ref{fig:PZTdata}b which shows $I_{sd}$ vs $V_{gate}$ curves for screening fractions ranging from 0\% to 75\% and is in excellent agreement with our experimental data. We note that the simulations agree with our data for $\mu > 0$ and $\delta \geq 0$, which is consistent with the extrinsic p-doping typically observed in graphene devices fabricated on PbZr$_{0.2}$Ti$_{0.8}$O$_3$\cite{Baeumer2013}.

The substrate-selectivity of the doping in our devices suggests that the ambient dopants are polar H$_2$O molecules. Polar surface adsorbates have been shown to dramatically affect the electronic properties of complex oxide systems\cite{Xie2011} as well as graphene\cite{Schedin2007, Lohmann2009,Lu2014}. The H$_2$O doping is substrate-selective because of the unique ferroelectric nature of the PbZr$_{0.2}$Ti$_{0.8}$O$_3$ substrate. Previous work\cite{Leenaerts2007,Leenaerts2009} has established the importance of the orbital structure of the adsorbate in determining the most energetically favorable orientation. For standard graphene devices on SiO$_2$ substrates, these energetics favor a uniform polarization throughout the range of applicable gate voltages. In the devices described here, it is likely that the remnant polarization of the ferroelectric substrate sufficiently alters the orbital structure of the adsorbed H$_2$O molecules to destroy this stability, and thus create less-polarized regions. This hypothesis is in agreement with data collected from similar devices fabricated on non-ferroelectric substrates; such devices show none of the characteristic Dirac point splitting associated with p-n junctions.

\begin{figure}
  \captionsetup[subfigure]{labelformat=empty}
  \centering
  \subfloat[]{\includegraphics[width=.22\textwidth]{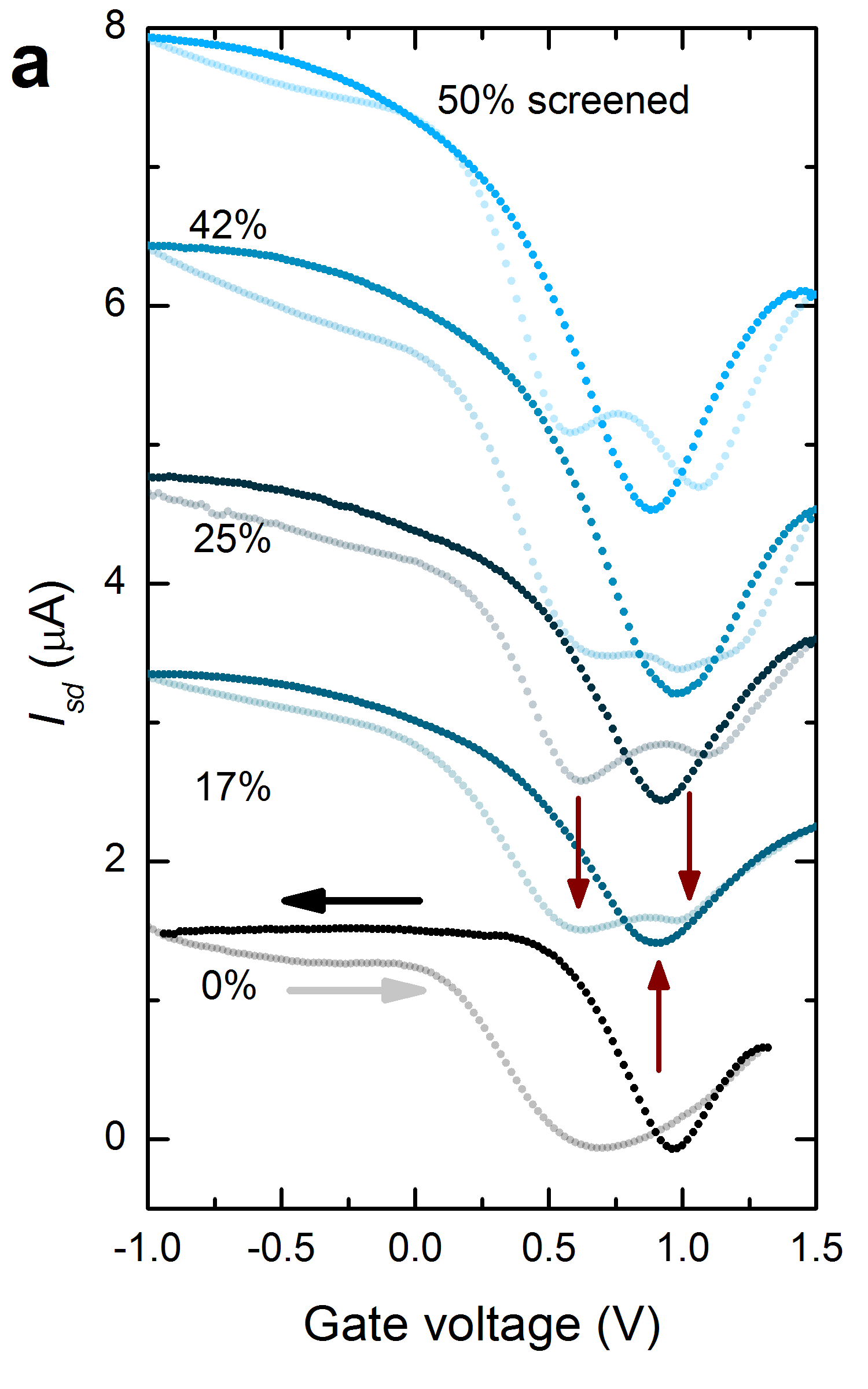}}\quad
  \subfloat[]{\includegraphics[width=.22\textwidth]{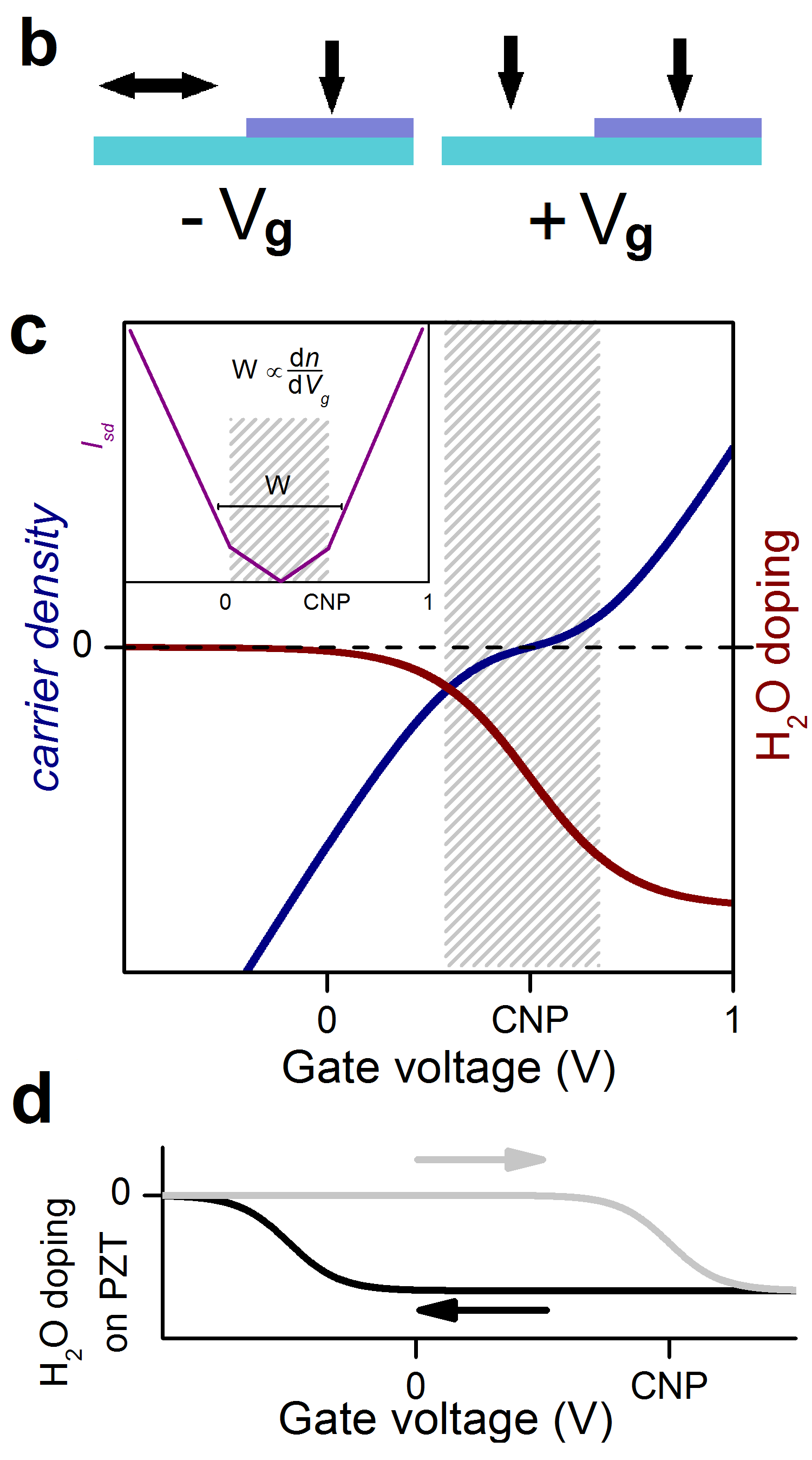}}\quad
  \caption{
   (a) Offset $I_{sd}$ vs $V_{gate}$ curves for forward (light) and reverse (dark) gate voltage sweeps. In all non-control devices, forward sweeps show split Dirac points, but reverse sweeps show only a single Dirac point. Representative minima locations are indicated by the vertical arrows.
   (b) H$_2$O polarization by device region, as prepared by positive and negative gate voltages. For negative gate voltages, H$_2$O on the PbZr$_{0.2}$Ti$_{0.8}$O$_3$ region is unpolarized, and H$_2$O on the SiO$_2$ region is polarized. For positive gate voltages both regions have a net polarization.
   (c) Simulated carrier density vs. applied gate voltage. The onset of H$_2$O polarization (indicated by grey shading) decreases the slope of the carrier density vs. gate voltage curve for graphene on PbZr$_{0.2}$Ti$_{0.8}$O$_3$. Inset: simulated conductance vs. applied gate voltage; the slope of carrier density vs. gate voltage at the CNP determines the width of the conductance minimum.
   (d) A schematic illustration of the H$_2$O doping hysteresis with applied gate voltage; arrows indicate the direction of the gate voltage sweep. The application of a large negative gate voltage destroys the H$_2$O polarization while a large positive gate voltage establishes a net polarization.
  }
  \label{fig:doping_dynamics}
\end{figure}

The gate-voltage dependence of the H$_2$O polarization configurations on PbZr$_{0.2}$Ti$_{0.8}$O$_3$ vs SiO$_2$ leads to hysteresis in the devices. This can be seen in figure \ref{fig:doping_dynamics}a, which shows $I_{sd}$ vs $V_{gate}$ curves for forward and reverse gate sweeps performed on the same devices as measured in Fig. \ref{fig:PZTdata}a. A pronounced hysteresis between forward and reverse gate sweeps is apparent. As in Fig. \ref{fig:PZTdata}a, devices spanning a region of evaporated SiO$_2$ display two distinct minima during forward sweeps, while a control device having no SiO$_2$ displays a single minimum. However, all devices display a single minimum during reverse gate sweeps. We note that the gate voltages applied here remain below the coercive voltage of the PbZr$_{0.2}$Ti$_{0.8}$O$_3$ film, therefore ferroelectric switching is not a possible cause of the observed hysteresis.

\begin{figure}
  \captionsetup[subfigure]{labelformat=empty}
  \centering
  \subfloat[]{\includegraphics[width=.22\textwidth]{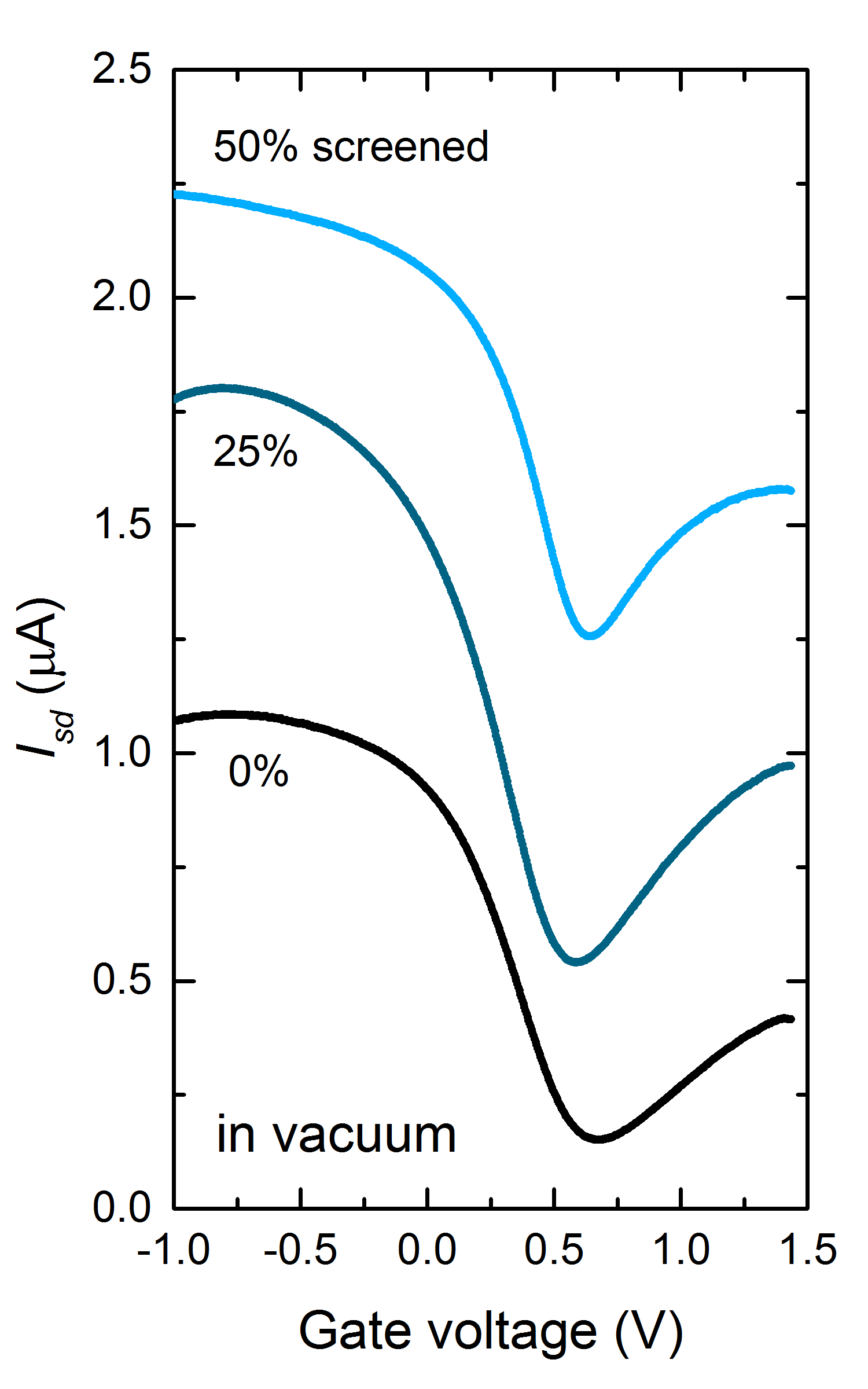}}\quad
  \caption{
   Offset $I_{sd}$ vs $V_{gate}$ curves for devices measured in vacuum. The Dirac point splitting effect disappears in vacuum, but can be recovered by leaving the device in ambient conditions overnight. From bottom to top the curves correspond to graphene channels which are 0, 25, and 50 percent screened.
  }
  \label{fig:PZTvacuum}
\end{figure}

In order to understand how the hysteresis is related to the different H$_2$O polarization configurations on the PbZr$_{0.2}$Ti$_{0.8}$O$_3$ as compared to the evaporated SiO$_2$, it is instructive to consider the gate voltages at which the various minima appear. For example, for the 17\% screened curve in Fig. \ref{fig:doping_dynamics}a the red arrows point out two minima on the forward sweep (at 0.6 V and 1.1V) and one minimum on the reverse sweep (at 0.9 V). These can be compared to the position of the Dirac point in vacuum at 0.6 V (see Fig. \ref{fig:PZTvacuum}). The minimum on the forward sweep at 0.6 V occurs at the same gate voltage as the vacuum Dirac point, implying that it corresponds to a region of the graphene without a net polarization in the adsorbed H$_2$O. The remaining minima occur at voltages larger than the Dirac point (0.9 V and 1.1 V) and thus correspond to regions of the graphene on which the adsorbed H$_2$O is polarized and produces p-doping. Polarized H$_2$O typically produces p-doping in graphene, though the precise mechanism is the subject of continuing research\cite{Leenaerts2007,Moser2008,Wang2010,Wehling2008,Sabio2008}. We identify the forward-sweep minimum at 0.6 V as corresponding to graphene on the non-screened PbZr$_{0.2}$Ti$_{0.8}$O$_3$ region. This is supported by the fact that all devices demonstrate a minimum at 0.6 V, independent of different SiO$_2$ screening fractions. In contrast, the minimum indicated by the rightmost arrow (e.g. at 1.1 V for the 17\% screened device) corresponds to graphene on the evaporated SiO$_2$ region, as evidenced by its evolution with increasing SiO$_2$ screening fraction.

We conclude that the application of a negative gate voltage destroys the net polarization of adsorbed H$_2$O on PbZr$_{0.2}$Ti$_{0.8}$O$_3$, but preserves a net polarization on the SiO$_2$-screened regions. This creates different local doping levels and thus a p-n junction. A positive gate voltage establishes a net polarization in both regions, and thus a uniform channel with no p-n junction, as depicted in Figure \ref{fig:doping_dynamics}b. This interpretation is further corroborated by the different widths of the forward and reverse minima for the control (0\% screened) device, as evident in Fig. \ref{fig:doping_dynamics}a. This difference can be explained by considering that conductance is linear with carrier density\cite{Hwang2007}, so the width of the conductance minimum at the Dirac point is determined by the slope of the carrier density vs. gate voltage curve. Typically the slope is constant, determined by the gate capacitance. However, for our devices the onset of H$_2$O dipole doping introduces a nonlinearity in the regime where the adsorbed H$_2$O transitions from unpolarized to polarized; this is shown schematically in Fig. \ref{fig:doping_dynamics}c. The polarized H$_2$O in our devices p-dopes the graphene, so the onset of its doping contribution temporarily reduces the slope of the carrier density vs. gate voltage curve, broadening the conductance minimum. During reverse sweeps, the transition from polarized to unpolarized H$_2$O occurs far from the CNP, so the width of the conductance minimum is unaffected. The hysteresis in H$_2$O polarization is illustrated schematically in Figure \ref{fig:doping_dynamics}d. Experimentally, the polarization hysteresis displays a dependence on both the magnitude of the applied gate voltage and the duration of its application, which prevents an exact determination of the gate voltages required to establish or destroy the H$_2$O polarization.

The hysteresis of the H$_2$O polarization on PbZr$_{0.2}$Ti$_{0.8}$O$_3$ substrates adds an `on/off' switching element to the p-n behavior. Specifically, we can selectively transition the device into and out of the p-n junction configuration through the application of large positive and negative gate voltages. The initial application of a large positive gate voltages establishes a uniform polarization across the device, yielding a unipolar conducting channel, while a large negative gate voltage destabilizes the polarization on regions supported by PbZr$_{0.2}$Ti$_{0.8}$O$_3$. In the latter case the different H$_2$O polarizations create separate locally doped regions, and thus a p-n junction. This `on/off' switching is different than the standard gate induced switching observed in p-n junctions, for example from p-n to p$^+$-p. By comparison, in our devices the same applied gate voltage can generate either a p-n junction or a uniformly doped channel, depending on the H$_2$O polarization condition.

The ferroelectric nature of the PbZr$_{0.2}$Ti$_{0.8}$O$_3$ substrate might suggest that the residual electric field from the substrate polarization dopes the regions of graphene in direct contact with the substrate\cite{Baeumer2013,Zheng2010}, but has less effect in the graphene regions screened by evaporated SiO$_2$. However, this explanation is precluded by several further experimental observations. First, the Dirac point splitting effect disappears when the devices are measured in vacuum. Figure \ref{fig:PZTvacuum} shows $I_{sd}$ vs $V_{gate}$ curves for the same devices measured at $5\times10^{-6}$ Torr but otherwise under conditions identical to those of Figure \ref{fig:PZTdata}a. All devices show a single minimum, independent of evaporated SiO$_2$ width or gate sweep rate. Leaving the devices in ambient conditions overnight recovers the splitting effect. The observed behavior is consistent with an ambient dopant mechanism, i.e., the substrate-selective formation of a self-assembled layer of dopant molecules. In vacuum, dopant molecules desorb from the surface leaving all regions of the graphene identically doped. Leaving the device in ambient conditions allows the dopant layer to reassemble, thereby re-establishing the separately doped regions.

The absence of Dirac point splitting in vacuum measurements also eliminates differences in gate capacitance as a dominant source of the splitting effect. The screened and non-screened regions of the device have different gate thicknesses and dielectric constants, which might suggest that the application of the same gate voltage would generate different doping levels in each region, and hence the transport behavior we observe. However, any capacitative differences between the regions are static, depending only on the geometry of the device, while the Dirac point splitting effect is dynamic, disappearing in vacuum. Capacitance-based explanations are further ruled out by the nearly identical Dirac point locations observed in all devices under vacuum. In particular, the similarity of data under vacuum from the control device (having no evaporated SiO$_2$), and from the devices which do span an evaporated SiO$_2$ region confirms that gate capacitance differences between the two regions are not the primary cause of the Dirac point splitting effect.

In summary, we have fabricated a controllable p-n junction in graphene on a ferroelectric substrate. We employ simple substrate modifications to define local doping regions, where the doping is accomplished through the substrate-selective formation of a self-assembled layer of ambient doping molecules. Alternative explanations for the local doping effect are ruled out, and the dynamics of the ambient doping suggest that it is due to polar H$_2$O molecules. Finally, the PbZr$_{0.2}$Ti$_{0.8}$O$_3$ substrate creates a hysteresis in the ambient doping effect which can be used to controllably bias the device into and out of a p-n junction configuration, using a single, universal backgate.




\textit{Acknowledgement.} J.H.H., R.X., S.R., and M.S. acknowledge support from the National Science Foundation and the Nanoelectronics Research Initiative under NSF-NEB grant DMR-1124696.
S.P. acknowledges support from the Army Research Office under grant W911NF-14-1-0104. N.M. and L.W.M. acknowledge support from the National Science Foundation under grant ENG-1434147. This work was carried out in part in the Frederick Seitz Materials Research Laboratory Central Facilities at the University of Illinois.

\bibliography{pztpaper-arXiv}

\end{document}